\documentclass[sigconf]{acmart}
\AtBeginDocument{%
  }

\copyrightyear{2026}
\acmYear{2026}
\setcopyright{cc}
\setcctype{by}
\acmConference[KDD '26]{Proceedings of the 32nd ACM SIGKDD Conference on Knowledge Discovery and Data Mining V.2}{August 09--13, 2026}{Jeju Island, Republic of Korea}
\acmBooktitle{Proceedings of the 32nd ACM SIGKDD Conference on Knowledge Discovery and Data Mining V.2 (KDD '26), August 09--13, 2026, Jeju Island, Republic of Korea}
\acmDOI{10.1145/3770855.3818429}
\acmISBN{979-8-4007-2259-2/2026/08}

\usepackage{balance}
\usepackage{multirow}

\begin{document}

\title{HGenPush: A Heterogeneous Generative Recommendation Architecture for Industrial Push Notification Systems}


\author{Xiao Liang}
\authornote{Both authors contributed equally to this research.}
\affiliation{%
  \institution{Kuaishou Technology}
  \city{Beijing}
  \country{China}}
\email{liangxiao@kuaishou.com}

\author{Jiali Feng}
\authornotemark[1]
\affiliation{%
  \institution{Kuaishou Technology}
  \city{Beijing}
  \country{China}}
\email{fengjiali05@kuaishou.com}

\author{Xin Feng}
\affiliation{%
  \institution{Kuaishou Technology}
  \city{Beijing}
  \country{China}}
\email{fengxin05@kuaishou.com}

\author{Yiqing Wang}
\affiliation{%
  \institution{Kuaishou Technology}
  \city{Beijing}
  \country{China}}
\email{wangyiqing05@kuaishou.com}

\author{Baolin Ye}
\affiliation{%
  \institution{Kuaishou Technology}
  \city{Beijing}
  \country{China}}
\email{yebaolin@kuaishou.com}

\author{Siyao Feng}
\affiliation{%
  \institution{Kuaishou Technology}
  \city{Beijing}
  \country{China}}
\email{fengsiyao@kuaishou.com}

\author{Zhihui Deng}
\affiliation{%
  \institution{Kuaishou Technology}
  \city{Beijing}
  \country{China}}
\email{dengzhihui@kuaishou.com}

\author{Cunyi Zhang}
\affiliation{%
  \institution{Kuaishou Technology}
  \city{Beijing}
  \country{China}}
\email{zhangcunyi@kuaishou.com}

\author{Huajin Sun}
\affiliation{%
  \institution{Kuaishou Technology}
  \city{Beijing}
  \country{China}}
\email{sunhuajin@kuaishou.com}

\author{Xuanping Li}
\authornote{Corresponding author.}
\affiliation{%
  \institution{Kuaishou Technology}
  \city{Beijing}
  \country{China}}
\email{lixuanping@kuaishou.com}

\author{Kaiqiao Zhan}
\affiliation{%
  \institution{Kuaishou Technology}
  \city{Beijing}
  \country{China}}
\email{zhankaiqiao@kuaishou.com}

\author{Yanan Niu}
\affiliation{%
  \institution{Kuaishou Technology}
  \city{Beijing}
  \country{China}}
\email{niuyanan@kuaishou.com}

\author{Kun Gai}
\affiliation{%
  \institution{Kuaishou Technology}
  \city{Beijing}
  \country{China}}
\email{gai.kun@qq.com}

\renewcommand{\shortauthors}{Xiao Liang et al.}

\begin{abstract}
  With the explosive growth of content platforms, recommendation systems need to better satisfy user demands to enhance user satisfaction and retention. Taking short-video platforms as an example, users not only seek high-quality content but also trusted authors. Although generative recommendation systems have achieved breakthroughs in recent years, existing methods primarily generate single-type recommendation content and typically employ the inefficient autoregressive paradigm to generate semantic IDs. In this paper, we propose an end-to-end heterogeneous generative recommendation architecture called HGenPush. First, we design a hybrid user behavior understanding module that integrates multi-scenario and multi-perspective behaviors to capture precise user interest. Then, we design a dual-branch heterogeneous generative recommendation module that integrates video recommendation and author recommendation within a unified framework. In addition, to improve generation efficiency, we design a lightweight multi-token prediction method that discards the autoregressive paradigm. Finally, we design a user consumption preference alignment module, which leverages user feedback as reward signals to guide the model toward generating higher-quality content, thereby enhancing user experience and engagement. Through these designs, HGenPush simultaneously fulfills users' demands for high-quality content and trusted authors. We have deployed HGenPush on the push notification system of Kuaishou, a large-scale short-video platform, achieving a significant 0.181\% increase in daily active users.
\end{abstract}



\maketitle

\section{Introduction}
\label{sec:introduction}

In short-video platforms, users are drawn not only to remarkable video content but also to the personalities of authors. The similar phenomenon also exists in other content platforms. For instance, news readers follow both breaking events and trusted media outlets, while e-commerce consumers not only purchase desired products but also explore new products from trusted merchants. Therefore, building a heterogeneous recommendation system that integrates both high-quality content and trusted source recommendations has significant practical value.

In recent years, generative large language models have revolutionized recommendation systems. Many studies \cite{li2023text, li2024calrec,zhai2024actions, guo2025onesug} demonstrate that Transformer-based sequence models can provide personalized recommendation to users in an end-to-end way. These methods learn user interest from user behavior sequences and output items that users may be interested in, without relying on traditional multi-stage filtering pipeline. As a pioneering method, TIGER \cite{rajput2023recommender} introduces RQ-VAE to encode item content information into semantic IDs, enabling knowledge sharing among similar items. OneRec \cite{deng2025onerec} proposes a generative recommendation approach that leverages reward models to match user preference. Subsequently, methods such as OneSug \cite{guo2025onesug}, EGA \cite{zheng2025ega}, and OneLoc \cite{wei2025oneloc} introduce adaptive enhancements for scenario-specific requirements, further advancing the implementation of generative recommendation paradigm in industrial systems.

Despite demonstrating significant potential, existing generative recommendation methods face two critical challenges. First, current methods primarily focus on homogeneous content (i.e., items of a single type), struggling to jointly recommend heterogeneous content such as videos and authors. Second, mainstream autoregressive semantic IDs generation limits inference speed and demands substantial computational resources, restricting the deployment of generative recommendation models in production systems.

To address these challenges, we propose HGenPush, an end-to-end heterogeneous generative recommendation architecture for short-video push notification systems. As illustrated in Figure~\ref{fig:main}, the architecture adopts a decoder-only structure that takes user behavior sequences and important static features as inputs, then employs a carefully designed heterogeneous generative recommendation module to generate videos and authors aligned with user interest. HGenPush comprises three core modules: (1) Hybrid User Behavior Understanding Module. Considering the synergistic effects between push and feed recommendation, we design a multi-perspective user behavior fusion solution. We integrate users' long-term and short-term feed behaviors to establish foundational user interest, incorporate the push click sequence to model push-specific behavior patterns, and introduce the push send sequence to ensure recommendation validity. (2) Heterogeneous Generative Recommendation Module. To efficiently generate semantic IDs for videos and authors, we propose Chained Multi-Token Prediction (Chained-MTP), which replaces autoregressive computation by using user interest representation as stable context and modeling semantic dependencies through cumulative embeddings of preceding semantic IDs. Then, in the video recommendation branch, we utilize semantic IDs encoded from multimodal video information to represent videos and execute generative recommendation processes based on this representation, thereby enabling the matching of video content with user interest. The essence of author recommendation is identifying users' willingness to consistently follow an author based on their behaviors. Therefore, the author recommendation branch learns author encoding that captures user behavior preference through representation alignment, and discretizes it into semantic IDs for generation. (3) User Consumption Preference Alignment (UCPA) Module. Since clicks primarily reflect initial interest rather than content value or satisfaction, we follow the Reinforcement Learning with Verifiable Rewards (RLVR) paradigm \cite{lambert2024tulu}, utilizing reward signals generated from real user feedback to continuously optimize the model. Inspired by CISPO \cite{chen2025minimax} and GSPO \cite{zheng2025group}, we propose Group Sequence Importance Sampling Policy Optimization (GSISPO), which employs sequence-level rewards and importance sampling weights to prevent semantic fragmentation and uses importance sampling weight clipping to preserve gradient signals from niche but high-value behaviors.

Overall, the main contributions of this paper are as follows:
\begin{itemize}
\item We systematically summarized the two types of user demands during content consumption and propose HGenPush, a novel end-to-end heterogeneous generative recommendation architecture that jointly recommends videos and authors.
\item We design a lightweight multi-token prediction method called Chained-MTP. It avoids interest drift through user interest representation and preserves semantic dependencies via cumulative embeddings of semantic IDs, while significantly improving inference speed.
\item We design a user consumption preference alignment module that leverages real user feedback to optimize the model toward generating high-quality content, enhancing user experience and retention.
\item We have successfully deployed HGenPush on Kuaishou's push notification system. Extensive experiments demonstrate its effectiveness and efficiency in heterogeneous recommendation.
\end{itemize}

\section{Related Work}
\label{sec:relatedWork}

\subsection{Generative Recommendation}
\label{sec:generativeRecommendation}

In recent years, generative recommendation \cite{wang2024enhanced, wang2025generative}, as an emerging paradigm, has been profoundly transforming the methodological framework of recommendation technologies. Generative recommendation reframes the recommendation task as a sequence generation problem, employing Transformer-based architectures to deeply model user behaviors and directly generate items matching user interest in an end-to-end manner. This paradigm has opened a new pathway for enhancing the performance of recommendation systems. TIGER \cite{rajput2023recommender} innovatively encodes items into meaningful semantic ID tuples via RQ-VAE to enable knowledge sharing among similar items. It then decodes target item's semantic IDs through autoregressive decoding, thus achieving end-to-end recommendation. Considering the semantic gap between large language models and recommendation systems, LC-REC \cite{zheng2024adapting} integrates language and collaborative semantics into large language models for recommendation. OneRec \cite{deng2025onerec} employs an encoder-decoder architecture to generate videos and enhances generation quality through a session-level preference alignment strategy. EGA \cite{zheng2025ega} integrates user interest modeling, ad slot and creative generation, position allocation, and payment optimization within a single model, enabling the practical application of generative architectures in industrial advertising system. While these generative recommendation methods demonstrate significant potential, they primarily focus on single-type recommendation, struggling to simultaneously meet heterogeneous recommendation objectives.

\subsection{Reinforcement Learning}
\label{sec:reinforcementLearning}

For generative recommendation systems, relying solely on sequence generation task is insufficient to satisfy diverse recommendation requirements or business constraints. Therefore, most methods utilize reinforcement learning to guide model iteration, thereby effectively aligning recommendation results with practical needs. GRPO \cite{shao2024deepseekmath} calculates relative ranking advantages within groups for multiple candidate outputs from generative recommendation models, updating policies using the PPO-style \cite{schulman2017proximal} objective function without requiring additional critic model. GSPO \cite{zheng2025group} employs sequence-level optimization to avoid the logical fragmentation problem caused by traditional token-level optimization. CISPO \cite{chen2025minimax} uses clipped importance sampling weight and gradient clipping to address contribution loss due to gradient blocking.

\section{Methodology}
\label{sec:methodology}

\begin{figure*}[htbp]
  \centering 
  \includegraphics[width=\textwidth]{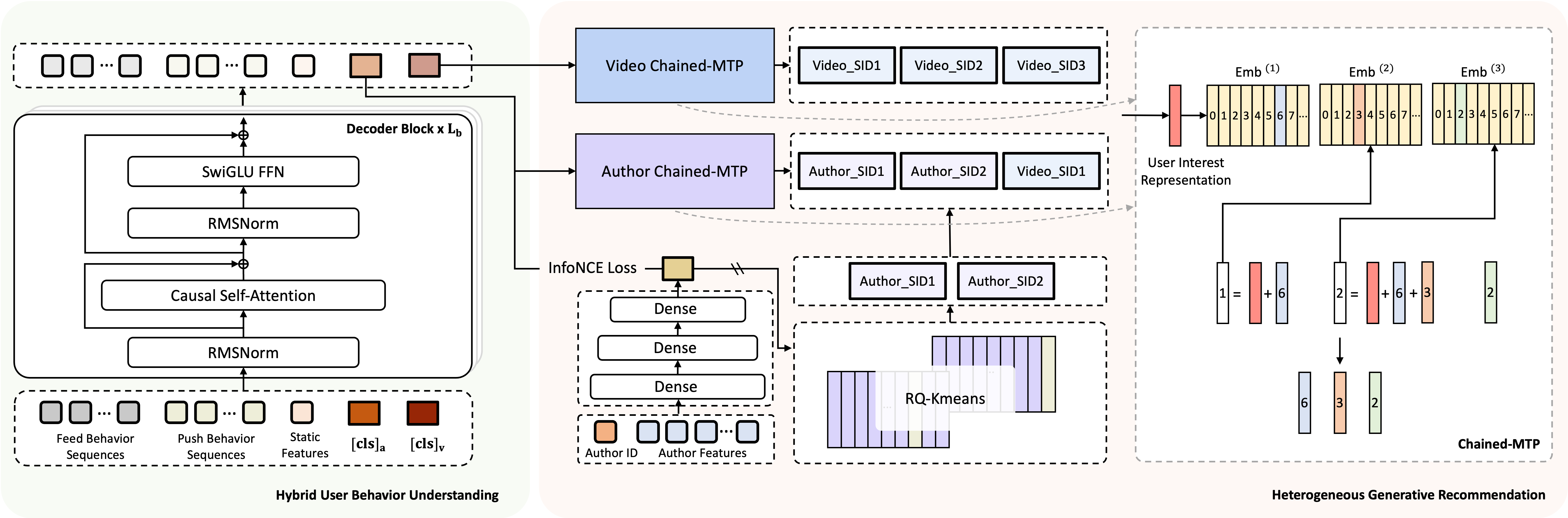}  
  \caption{The overall architecture of HGenPush, where "\textbackslash\textbackslash" denotes stop gradient.}
  \label{fig:main}
\end{figure*}
In this section, we introduce the overall architecture of HGenPush. Specifically, Section~\ref{sec:userBehaviorUnderstanding} details the modeling process of hybrid user behavior understanding module. Section~\ref{sec:heterogeneousGenerativeRecommendation} describes the specific workflow of different branches within the heterogeneous generative recommendation module. Section~\ref{sec:UCPA} explains how to further enhance the quality of model-generated results through reinforcement learning. Section~\ref{sec:trainingAndInference} presents the joint training loss and inference strategy. The overall architecture of HGenPush is illustrated in Figure~\ref{fig:main}.

\subsection{Hybrid User Behavior Understanding}
\label{sec:userBehaviorUnderstanding}

Precise user interest modeling is the core challenge for recommendation systems. While push notification recommendation and feed recommendation differ in user interaction formats, they exhibit synergistic effects in user behaviors. For example, users who frequently watch gaming videos or deeply browse an author's feed content are more likely to return when receiving relevant push notifications. This cross-scenario behavior correlation can not only enhance user experience but also deepen users' recognition of the platform's value. Considering this characteristic, we introduce multi-scenario and multi-perspective behaviors into our recommendation system. Simultaneously, we also introduce important static features (such as user gender and age) as foundational user information. Finally, we generate the user interest representation through a decoder-only architecture.

\subsubsection{Feature Representation}
\label{sec:featureRepresentation}

We denote user behavior sequences as $S$, which includes both feed sequences $S_{feed}$ and push notification sequences $S_{push}$. $S_{feed}$ integrates long-term ($S_{long}$) and short-term ($S_{short}$) behavior sequences to model foundational user interest. $S_{push}$ incorporates the push click sequence ($S_{click}$) capturing users' push-specific behavior patterns and the push send sequence ($S_{send}$) ensuring recommendation validity. Each video $v$ is represented by semantic IDs generated through its multimodal embedding, and the generation method is consistent with OneRec \cite{deng2025onerec}. We define the semantic IDs as $s_v=(s_v^1, s_v^2, \cdots, s_v^{L_v})$, where $L_v$ denotes the number of codebooks and the size of each codebook is $N$.

First, we transform each semantic token into a dense embedding and concatenate them to obtain video semantic representation $e_{semantic}^v=[e_{s_v^1}; e_{s_v^2};\cdots;e_{s_v^{L_v}}]$. Then, we introduce additional features to further distinguish the impact differences of videos.

For each video in the long-term behavior sequence $S_{long}$, we introduce author identifier (aid), topic tag (tag), and playtime (playtime). And we map the concatenated result of feature embeddings to a unified dimension $d_h$ as the final representation for the video. Therefore, the representation of the $i$-th video $v_{long}^i$ in $S_{long}$ is defined as follows:
\begin{equation}
    h_{v_{long}}^i=\sigma(\operatorname{Dense}([e_{semantic}^{v_{long}^i};e_{aid}^{v_{long}^i};e_{tag}^{v_{long}^i};e_{playtime}^{v_{long}^i}])),
\end{equation}
where $\sigma(\cdot)$ denotes the activation function.

The overall representation of long-term behavior sequence is denoted as:
\begin{equation}
    h_{long}=\{h_{v_{long}}^1, h_{v_{long}}^2,\cdots,h_{v_{long}}^{L_{long}}\},
\end{equation}
where $L_{long}$ is the length of $S_{long}$ and $h_{long}\in \mathcal{R}^{L_{long} \times d_h}$. This representation is primarily used to construct stable and comprehensive user interest preference. 

The short-term sequence $S_{short}$ consists of videos that the user actively engaged with in recent times, reflecting changing user interest. Its representation generation process aligns with the one mentioned above:
\begin{equation}
    h_{v_{short}}^i=\sigma(\operatorname{Dense}([e_{semantic}^{v_{short}^i};e_{aid}^{v_{short}^i};e_{tag}^{v_{short}^i};e_{playtime}^{v_{short}^i}])),
\end{equation}
\begin{equation}
    h_{short}=\{h_{v_{short}}^1, h_{v_{short}}^2,\cdots,h_{v_{short}}^{L_{short}}\}.
\end{equation}
Here, $L_{short}$ denotes the length of $S_{short}$ and $h_{short}\in \mathcal{R}^{L_{short} \times d_h}$.

User behaviors in the push notification scenario directly reflect their preferences and habits toward notifications. Therefore, we introduce user push click sequence. Specifically, we add a notification type (type) feature to learn which content types the user is more likely to click. The representation generation process for push click sequence is as follows:
\begin{equation}
    h_{v_{click}}^i=\sigma(\operatorname{Dense}([e_{semantic}^{v_{click}^i};e_{aid}^{v_{click}^i};e_{type}^{v_{click}^i}])),
\end{equation}
\begin{equation}
    h_{click}=\{h_{v_{click}}^1, h_{v_{click}}^2,\cdots,h_{v_{click}}^{L_{click}}\},
\end{equation}
where $L_{click}$ denotes the length of $S_{click}$ and $h_{click}\in \mathcal{R}^{L_{click} \times d_h}$.

In addition, we introduce push send sequence to construct contextual information, thereby enhancing the rationality of generation:
\begin{equation}
    h_{v_{send}}^i=\sigma(\operatorname{Dense}([e_{semantic}^{v_{send}^i};e_{aid}^{v_{send}^i};e_{type}^{v_{send}^i}])),
\end{equation}
\begin{equation}
    h_{send}=\{h_{v_{send}}^1, h_{v_{send}}^2,\cdots,h_{v_{send}}^{L_{send}}\},
\end{equation}
where $L_{send}$ denotes the length of $S_{send}$ and $h_{send}\in \mathcal{R}^{L_{send} \times d_h}$.

User static features serve as the "stabilizer" for user interest modeling. They can provide reliable signals when user behaviors are sparse. Therefore, we introduce supplementary features such as user age (age) and gender (gender). Then, we concatenate their embeddings and map them to a unified dimension $d_h$:
\begin{equation}
    h_{static}=\sigma(\operatorname{Dense}([e_{age};e_{gender};\cdots])).
\end{equation}

Finally, we concatenate all representations to form the complete input representation $h_f$:
\begin{equation}
    h_{f}=[h_{long};h_{short};h_{click};h_{send};h_{static};h_{[\text{cls}]_a};h_{[\text{cls}]_v}].
\end{equation}

Notably, we append special tokens $h_{[\text{cls}]_a}$ and $h_{[\text{cls}]_v}$ at the end of the sequence features and static features, which are used for the author recommendation branch and video recommendation branch respectively. These two special tokens can summarize user interest and provide initial context for the decoder.

\subsubsection{Interest Understanding}
\label{sec:interestUnderstanding}

To efficiently fuse user static features with dynamic behavior sequences for precise personalized recommendations, we design a decoder-only architecture for the user interest understanding module.

As shown in Figure~\ref{fig:main}, this module stacks multiple blocks with identical structures. Each block primarily consists of a causal self-attention structure and a feed forward neural network, supplemented by residual networks to stabilize gradients. We define the input to the initial layer as $h^{(0)} = h_f$. The overall computational flow of this module is as follows:
\begin{equation}
    h_{attn}^{(l+1)}=\mathbf{CausalSelfAttn}(\mathbf{RMSNorm}(h^{(l)})),
\end{equation}
\begin{equation}
    h_{add}^{(l+1)}=h^{(l)}+h_{attn}^{(l+1)},
\end{equation}
\begin{equation}
    h^{(l+1)}=h_{add}^{(l+1)}+\mathbf{FFN_{SwiGLU}}(\mathbf{RMSNorm}(h_{add}^{(l+1)})).
\end{equation}

We utilize RMSNorm \cite{zhang2019root} to enhance training stability and employ SwiGLU FFNs \cite{shazeer2020glu} to further improve the expression ability of the model. By stacking $L_b$ layers, we can obtain the hidden representation ${h^{(L_b)}}$ of the final block. Then, we normalize it to obtain the final output $h^{\prime}=\{{h_{v_{long}}^{1}}^{\prime},{h_{v_{long}}^2}^{\prime},\cdots,h_{[\text{cls}]_a}^{\prime},h_{[\text{cls}]_v}^{\prime}\}$. Here, $h_{[\text{cls}]_a}^{\prime}$ and $h_{[\text{cls}]_v}^{\prime}$ serve as sequence endpoints with a global perspective. They not only encode static features but also comprehensively integrate all preceding behaviors, forming temporal-aware interest representations. Therefore, we feed $h_{[\text{cls}]_a}^{\prime}$ and $h_{[\text{cls}]_v}^{\prime}$ into subsequent heterogeneous generative recommendation module.

\subsection{Heterogeneous Generative Recommendation}
\label{sec:heterogeneousGenerativeRecommendation}

The heterogeneous generative recommendation module optimizes video and author recommendation branches synchronously. In the video recommendation branch, we propose a novel multi-token prediction method called Chained-MTP to generate semantic IDs efficiently. In the author recommendation branch, we design an end-to-end generative architecture with behavior-based representation alignment, which simultaneously optimizes preference-aware author encoding and trusted author generation.

\subsubsection{Video Recommendation Branch}
\label{sec:videoRecommendationBranch}

The video recommendation branch aims to generate results matching user interest from a content perspective. To achieve this, we leverage the global interest representation $h_{[\text{cls}]_v}^{\prime}$ learned from user behavior sequences for generation, which encodes both video semantic features and temporal evolution patterns.

\paragraph{\textbf{Chained Multi-token Prediction (Chained-MTP)}} Mainstream generative recommendation methods typically rely on multiple forward passes, resulting in inefficient inference. This limitation hinders the deployment of generative recommendation systems in resource-constrained scenarios. Therefore, inspired by RQ-Kmeans \cite{deng2025onerec, luo2025qarm}, we propose a novel chained multi-token prediction method, which can maintain semantic dependencies and avoid the multi-round decoding. Specifically, defining the semantic IDs of target video $v_t$ as $s_{v_t} = (s_{v_t}^1, s_{v_t}^2, \cdots, s_{v_t}^{L_v})$, we utilize $h_{[\text{cls}]_v}^{\prime}$ as the stable interest anchor and compute the cumulative embedding of the preceding semantic IDs at each step to model conditional dependencies across semantic levels:
\begin{equation}
    p_{s_{v_t}^{k}}=\mathbf{softmax}(\mathbf{FFN}^{({k})}(h_{[\text{cls}]_v}^{\prime}+\sum_{j<k}\mathbf{Emb}^{(j)}(s_{v_t}^j)))_{s_{v_t}^{k}},
\end{equation}
where $p_{s_{v_t}^{k}}$ denotes the generation probability of the $k$-th semantic token $s_{v_t}^{k}$, and $k \in \{1,2,\cdots,L_v\}$. As shown in the equation, we maintain a set of learnable embedding tables $\{\mathbf{Emb}^{(k)}\}_{k\in \{1,2,\cdots L_v\}}$ for hierarchical codebooks, where each $\mathbf{Emb}^{(k)}\in R^{N\times d_h}$ encodes the semantic tokens at level $k$, thereby characterizing semantic spaces at different granularities. It can be seen that we discard the complex autoregressive paradigm and significantly reduce computations through parallel feedforward network. In the training stage, we employ the cross-entropy loss function for optimization. Specifically, the loss function for the video recommendation branch is defined as follows:
\begin{equation}
    \mathcal{L}_{v-MTP} = -\frac{1}{B}\sum_{b=1}^{B}\sum_{k=1}^{L_v}\log(p_{s_{v_b}^k}),
\end{equation}
where $B$ denotes training batch size and $p_{s_{v_b}^k}$ represents the generation probability of the $k$-th semantic token for sample $b$. We construct training samples based on user click data in the push notification system.

\subsubsection{Author Recommendation Branch}
\label{sec:authorRecommendationBranch}

The author recommendation branch identifies authors likely to sustain user engagement and pushes their videos to users. In practice, user trust in authors is strongly reflected in their behaviors: even when two authors belong to similar categories, users may favor only one based on behavior preference that cannot be inferred from content alone. We therefore design the author recommendation branch as an end-to-end generative architecture with behavior-based representation alignment, which simultaneously learns preference-aware author encoding and trusted author generation.

\paragraph{\textbf{Behavior-based Representation Alignment}} Semantic IDs generated solely from author features struggle to distinguish user preferences among category-similar authors. To address this, we align author representation with user interest representation using user click data, thereby learning bidirectional audience information between users and authors to break through the limitation. First, we encode target author features (e.g., author identifier and multimodal embeddings) into latent embeddings $h_a^{\prime}$. Next, we project both user interest representation $h_{[\text{cls}]_a}^{\prime}$ and author representation $h_a^{\prime}$ into a unified feature space, yielding new representations $h_{u}$ and $h_a$. We then align these representations using the InfoNCE contrastive loss:
\begin{equation}
    \mathcal {L}_{align}=-\frac{1}{B}\sum_{b=1}^{B} \log\frac{\exp(\cos(h_{u_b},h_{a_b})/\tau)}{\sum_{b^{\prime}=1}^{B} \exp(\cos(h_{u_b},h_{a_{b^{\prime}}})/\tau)},
\end{equation}
where $B$ denotes training batch size, $\tau$ is the temperature coefficient, and $\cos (\cdot)$ calculates the cosine similarity between two input embeddings.

\paragraph{\textbf{Author Semantic IDs Construction}} To enable efficient trusted author generation, we discretize the aligned author embeddings into semantic IDs. Specifically, we employ the RQ-Kmeans algorithm for discretization and synchronously update the codebook during model training. Let $L_a$ denotes the number of author codebooks, $N$ denotes the size of each codebook, and $h_{a_t}$ denote the embedding representation of target author $a_t$. At the first layer, the initial residual is defined as $\boldsymbol{r}_{a_t}^1=h_{a_t}$. We obtain the index of the nearest centroid node embedding for layer $l$ through the following formula:
 \begin{equation}
     s_{a_t}^{l} = \arg\min_n \| \boldsymbol{r}_{a_t}^{l} - \boldsymbol{c}_{l}^n \|_2^2,
 \end{equation}
where $\boldsymbol{c}_{l}^n$ denotes the embedding of the $n$-th centroid node in the $l$-th codebook, with $l\in\{1,2,\cdots,L_a\}$. After obtaining $s_{a_t}^{l}$, the residual for the next layer is generated via $r_{a_t}^{l+1}=r_{a_t}^{l}-\boldsymbol{c}_l^{s_{a_t}^l}$.

Through iterative computation, we can obtain the author semantic IDs $s_{a_t}=(s_{a_t}^1, s_{a_t}^2,\cdots,s_{a_t}^{L_a})$. We employ the SimVQ strategy \cite{zhu2025addressing} to further stabilize the discretization process and prevent codebook collapse. Furthermore, we compute the Euclidean distance between the residual of each layer and its nearest centroid node embedding as the loss function to continuously update the codebook during training. The loss function formula is as follows:
\begin{equation}
    \mathcal {L}_{codebook}=\frac{1}{B}\sum_{b=1}^{B} \sum_{l=1}^{L_a} \|\operatorname{sg}(\boldsymbol{r}_{a_b}^{l}) - \boldsymbol{c}_l^{s_{a_b}^l}\|_2^2.
\end{equation}

\paragraph{\textbf{Mixed Semantic IDs Generation}} Since the author recommendation branch ultimately recommends videos from trusted authors, we propose mixed semantic IDs as the generation target. Specifically, we combine author semantic IDs and the first-level video semantic ID of target item to form $s_{m_t}=(s_{a_t}^1,s_{a_t}^2,\cdots,s_{a_t}^{L_a},s_{v_t}^1)$ with length $L_m=L_a+1$. We still apply Chained-MTP based on $h_{[\text{cls}]_a}^{\prime}$ for efficient generation. The loss function for the author recommendation branch is defined as follows:
\begin{equation}
    \mathcal{L}_{a-MTP} = -\frac{1}{B}\sum_{b=1}^{B}\sum_{x=1}^{L_m}\log(p_{s_{m_b}^x}).
\end{equation}

\subsection{User Consumption Preference Alignment}
\label{sec:UCPA}

While the model achieves stable recommendation capability through supervised training, relying solely on click data fails to capture content quality and user satisfaction, leading to clickbait recommendations. In fact, users' consumption behaviors within the application not only provide richer signals about their interest but also help the model identify the inherent value of content. Therefore, we design a user consumption preference alignment module following the Reinforcement Learning with Verifiable Rewards (RLVR) paradigm. The module utilizes post-click session feedback as rewards and optimizes the model via our proposed GSISPO algorithm.

\subsubsection{Reward Signals}
\label{sec:rewardSignals}

We define the $G$ videos consumed during the post-click session as $V=\{v_1,v_2,\cdots,v_G\}$. For each video $v_g$ in $V$, we derive its reward score $r_g$ by quantifying user consumption behavior according to fixed rules:
\begin{equation}
r_g =
\begin{cases}
r_g + 2 & \text{if } \text{label}_g \in \{\text{valid\_play}, \text{enter\_profile}, \text{comment\_stay}\} \\
r_g + 1 & \text{if } \text{label}_g \in \{\text{like}, \text{follow}, \text{share}, \text{comment}\} \\
r_g - 2 & \text{if } \text{label}_g \in \{\text{short\_play}\} \\
r_g - 5 & \text{if } \text{label}_g \in \{\text{dislike}\} \\
r_g - 10 & \text{if } \text{label}_g \in \{\text{report}\}
\end{cases},
\end{equation}
where $r_g$ is initialized to $0$ and accumulates rewards from all matched conditions. This reward structure incentivizes engaging content generation while discouraging low-quality or user-disliked videos through negative feedback.

\subsubsection{Group Sequence Importance Sampling Policy Optimization}
\label{sec:GSISPO}

Let $R=\{r_1,r_2,\cdots,r_G\}$ denote the rewards for the set of videos $V$. We first normalize all rewards to obtain each video's relative advantage within the group:
\begin{equation}
    \widehat{A}_g = \frac{r_g - \operatorname{mean}(r_1,r_2,\cdots,r_G)}{\operatorname{std}(r_1,r_2,\cdots,r_G)}.
\end{equation}

Chained-MTP generates the next semantic ID based on cumulative embedding of preceding results to ensure semantic coherence of the entire sequence. Therefore, we reference GSPO \cite{zheng2025group} to compute sequence-level importance sampling weight and generation probability, avoiding disruption to sequences' complete semantics and enhancing training stability. Moreover, since the relative advantage is specific to the video rather than individual semantic ID, sequence-level computation also aligns with our optimization objective. Simultaneously, we follow the CISPO \cite{chen2025minimax} algorithm by moving the clipping operation to the importance sampling weight, preserving gradient signals for niche yet high-value behaviors. Ultimately, the optimization objective is defined as follows:
\begin{equation}
    \mathcal{L}_{\text{GSISPO}}(\theta) = -\frac{1}{G} \sum_{g=1}^G \operatorname{sg} ( w_g(\theta)) \widehat{A}_g \operatorname{log}\pi_\theta(s_{v_g}|h_{[\text{cls}]_v}^{\prime}),
\end{equation}
where $s_{v_g}=\{s_{v_g}^1,s_{v_g}^2,\cdots,s_{v_g}^{L_v}\}$ denotes the semantic IDs of video $v_g$. $w_g(\theta) $ and $\pi_\theta(s_{v_g}|h_{[\text{cls}]_v}^{\prime})$ are the sequence-level importance sampling weight and model generation probability, respectively. The computation of $w_g(\theta) $ is as follows:
\begin{equation}
  \begin{split}
    w_g(\theta)^{\prime} &= \left( \frac{\pi_\theta(s_{v_g}|h_{[\text{cls}]_v}^{\prime})}{\pi_{\theta_{\text{old}}}(s_{v_g}|h_{[\text{cls}]_v}^{\prime})} \right)^{\frac{1}{L_v}} \\
                         & =\exp\left( \frac{1}{L_v} \sum_{k=1}^{L_v} \log \frac{\pi_\theta(s_{v_g}^k|h_{[\text{cls}]_v}^{\prime}, s_{v_g}^{<k})}{\pi_{\theta_{\text{old}}}(s_{v_g}^k|h_{[\text{cls}]_v}^{\prime}, s_{v_g}^{<k})} \right),
  \end{split}
\end{equation}
\begin{equation}
    w_g(\theta) = \operatorname{clip}\left(w_g(\theta)^{\prime}, 1-\epsilon, 1+\epsilon\right).
\end{equation}

Since $\mathcal{L}_{\text{GSISPO}}$ and $\mathcal{L}_{v-MTP}$ jointly guide model parameter updates, and $\mathcal{L}_{v-MTP}$ ensures model stability, our reinforcement learning optimization objective does not additionally introduce KL divergence.

\subsection{Training and Inference}
\label{sec:trainingAndInference}

Through the above modules, HGenPush achieves joint optimization of video recommendation, author recommendation, author codebook construction, and user consumption preference alignment. The overall training loss is defined as:
\begin{equation}
    \mathcal{L}=\mathcal{L}_{v-MTP}+\mathcal{L}_{a-MTP}+\mathcal {L}_{codebook}+\mathcal{L}_{\text{GSISPO}}.
\end{equation}

In the inference phase, both the video and author recommendation branches utilize Beam Search to generate multiple candidates. Our online system follows a standard cascaded recommendation architecture consisting of retrieval, pre-ranking, and ranking stages. Unlike traditional retrieval methods, candidates generated by HGenPush bypass the pre-ranking stage and are fed directly into the ranking model.

\section{Experiments}
\label{sec:experiments}

In this section, we will introduce the datasets, baseline methods, evaluation metrics, implementation details, as well as offline and online experimental analysis.

\subsection{Experimental Settings}
\label{sec:experimentsSettings}

\subsubsection{Datasets}
\label{sec:dataset}

To evaluate the effectiveness of HGenPush, we conduct experiments on Kuaishou's push notification system. This system records daily interactions from over 400 million active users, providing extensive and diverse real-world data for model training and evaluation. We continuously train our model using the online streaming data generated by the system. For offline evaluation, we load each model's checkpoint saved at the same time and evaluate them on the same dataset collected from the online system, ensuring comparability and reliability of experimental results.

\subsubsection{Baseline Methods}
\label{sec:baselineMethods}

We compare HGenPush with competitive traditional and generative recommendation models. The baseline methods are as follows:
\begin{itemize}
\item \textbf{SASRec} \cite{kang2018self} employs a unidirectional Transformer architecture, which effectively models the user sequence via self-attention mechanisms to predict the next item.
\item \textbf{TIGER} \cite{rajput2023recommender} proposes to replace item ID with hierarchical semantic IDs generated by RQ-VAE, and employs autoregressive generation techniques to implement sequence recommendations based on these IDs.
\end{itemize}

\subsubsection{Evaluation Metrics}
\label{sec:evaluationMetrics}

The model can generate multiple candidates at once through Beam Search. To evaluate model performance, we define HitRate@100 as our evaluation metric. Specifically, we calculate the proportion of samples whose true semantic IDs appear among the top 100 candidates:
\begin{equation}
    \text{HitRate@100}= \frac{1}{S}\sum_{i=1}^S I(s_i\in C_{100}^i),
\end{equation}
where $S$ denotes the number of samples, $s_i$ is the true semantic IDs for the $i$-th sample, and $C_{100}^i$ represents the top 100 candidates generated by the model for the $i$-th sample. $I(\cdot)$ is the binary indicator function, taking the value 1 when $s_i$ is in $C_{100}^i$ and 0 otherwise. This metric directly reflects the model's recommendation accuracy in real-world scenarios. A higher value indicates the model's greater ability to accurately capture user interest.

\subsubsection{Implementation Details}
\label{sec:implementationDetails}

HGenPush employs the Adam optimizer for training dense parameters and the AdamW optimizer for training sparse parameters, with both learning rates set to $1\times10^{-4}$. The weight decay is set to $1\times10^{-4}$, and the batch size is set to 64. Model training is performed on NVIDIA L20 GPUs. We set the number of decoder layers $L_b$ to 16, with a uniform representation dimension $d_h$ of 512. The number of video codebooks is 3 ($L_v=3$), and the size of each codebook is 8192 ($N=8192$). The number of author codebooks is 2 ($L_a=2$) and the size of each codebook is also 8192. In the reinforcement learning stage, we limit the number of videos per group to no more than 50.

\begin{table}[htbp]
  \centering
  \caption{Offline performance comparison between HGenPush and baseline methods. The best results are shown in bold.}
  \label{tab:offline}
  \begin{tabular}{lcc}
    \toprule
    \multirow{2}{*}{Model} & \multicolumn{2}{c}{HitRate@100} \\
    \cmidrule(lr){2-3}
    & Video Branch & Author Branch\\
    \midrule
    SASRec   & 0.3093 &0.3495 \\
    TIGER    & 0.3802 & 0.4677 \\
    HGenPush & \textbf{0.3915} & \textbf{0.4848} \\
  \bottomrule
\end{tabular}
\end{table}

\subsection{Offline Performance}
\label{sec:offlinePerformance}

As shown in Table~\ref{tab:offline}, both HGenPush and TIGER outperform SASRec, validating the power of semantic IDs and generative reasoning. Furthermore, HGenPush achieves optimal performance in both tasks due to its tailored designs. In the video recommendation branch, integrating the better-suited decoder-only architecture with Chained-MTP enables HGenPush to surpass the encoder-decoder-based TIGER in both effectiveness and efficiency (detailed analysis in Section~\ref{sec:architectureDesignValidation}). In the author recommendation branch, superior performance stems from two reasons: author semantic IDs can capture nuanced user preference for specific authors via behavior-based representation alignment, while mixed semantic IDs not only achieve author-level recommendation but also model which videos from preferred authors users favor, delivering more precise results.

\subsection{Architecture Design Validation}
\label{sec:architectureDesignValidation}

HGenPush employs a unified architecture for heterogeneous recommendation: both video and author branches share the same decoder-only backbone and Chained-MTP method for semantic IDs generation. To validate these core design choices, we conduct comparative experiments on the main model architecture and the multi-token prediction method. Since both branches adopt identical architectural components, we report HitRate@100 on the video recommendation task as a representative effectiveness metric (the author recommendation task exhibits consistent trends), while reporting model size and inference throughput (QPS) for the complete dual-branch model to reflect practical deployment.

\begin{table}
  \caption{Comparison of different main architectures.}
  \label{tab:mainArc}
  \begin{tabular}{cccc}
    \toprule
    Architecture&Model Size&HitRate@100&Loss\\
    \midrule
    Encoder-Decoder & 0.026B & 0.3400 & 11.73\\
    Decoder-Only & 0.016B & 0.3432 & 12.12\\
  \bottomrule
\end{tabular}
\end{table}

\subsubsection{Main Architecture}
\label{sec:mainArchitecture}

In this section, we compare decoder-only and encoder-decoder architectures. In the encoder-decoder variant, the encoder takes sequence features (excluding the push send sequence) and static features as input, while the decoder generates semantic IDs conditioned on the encoder's output and push send sequence. The offline metric comparison results are shown in Table~\ref{tab:mainArc}. As observed, the decoder-only architecture achieves comparable performance to the encoder-decoder structure but requires fewer parameters. This reduction directly translates to lower memory footprint and faster inference latency. Therefore, we ultimately adopt the decoder-only architecture for HGenPush.

\subsubsection{Multi-token Prediction}
\label{sec:mtp}

To validate the effectiveness of Chained-MTP, we compare it with two baselines:
\begin{itemize}
    \item \textbf{Parallel-MTP} predicts all semantic IDs independently without modeling semantic dependencies.
    \item \textbf{DeepSeek-MTP} \cite{liu2024deepseek} employs a hierarchical Transformer-based structure to generate semantic IDs.
\end{itemize}

The effectiveness and efficiency comparison results are shown in Table~\ref{tab:mtp}. As observed, Parallel-MTP suffers from significant performance degradation, demonstrating the necessity of modeling dependencies among semantic IDs. Compared with Parallel-MTP, DeepSeek-MTP substantially improves effectiveness through its Transformer-based cascaded structure, but at the cost of 47.09\% throughput reduction (1.27k vs 2.40k QPS).

In the trade-off between effectiveness and efficiency, Chained-MTP demonstrates superior comprehensive capability. It maintains competitive effectiveness metric to DeepSeek-MTP while delivering 33.86\% higher throughput. Chained-MTP's balance of effectiveness and efficiency stems from two design choices: (1) employing user interest representation as a stable global context baseline to prevent interest drift, and (2) computing cumulative embedding of preceding semantic IDs, which preserves semantic dependencies without the complex computations of Transformer.

\begin{table}
  \caption{Offline performance and inference efficiency comparison of different MTP methods.}
  \label{tab:mtp}
  \begin{tabular}{lcc}
    \toprule
    Method&HitRate@100&Max QPS per Device\\
    \midrule
    Parallel-MTP & 0.1150 & 2.40k \\
    DeepSeek-MTP & 0.3954 & 1.27k \\
    Chained-MTP & 0.3915 & 1.70k \\
  \bottomrule
\end{tabular}
\end{table}

\begin{figure}[h]
  \centering
  \includegraphics[width=\linewidth]{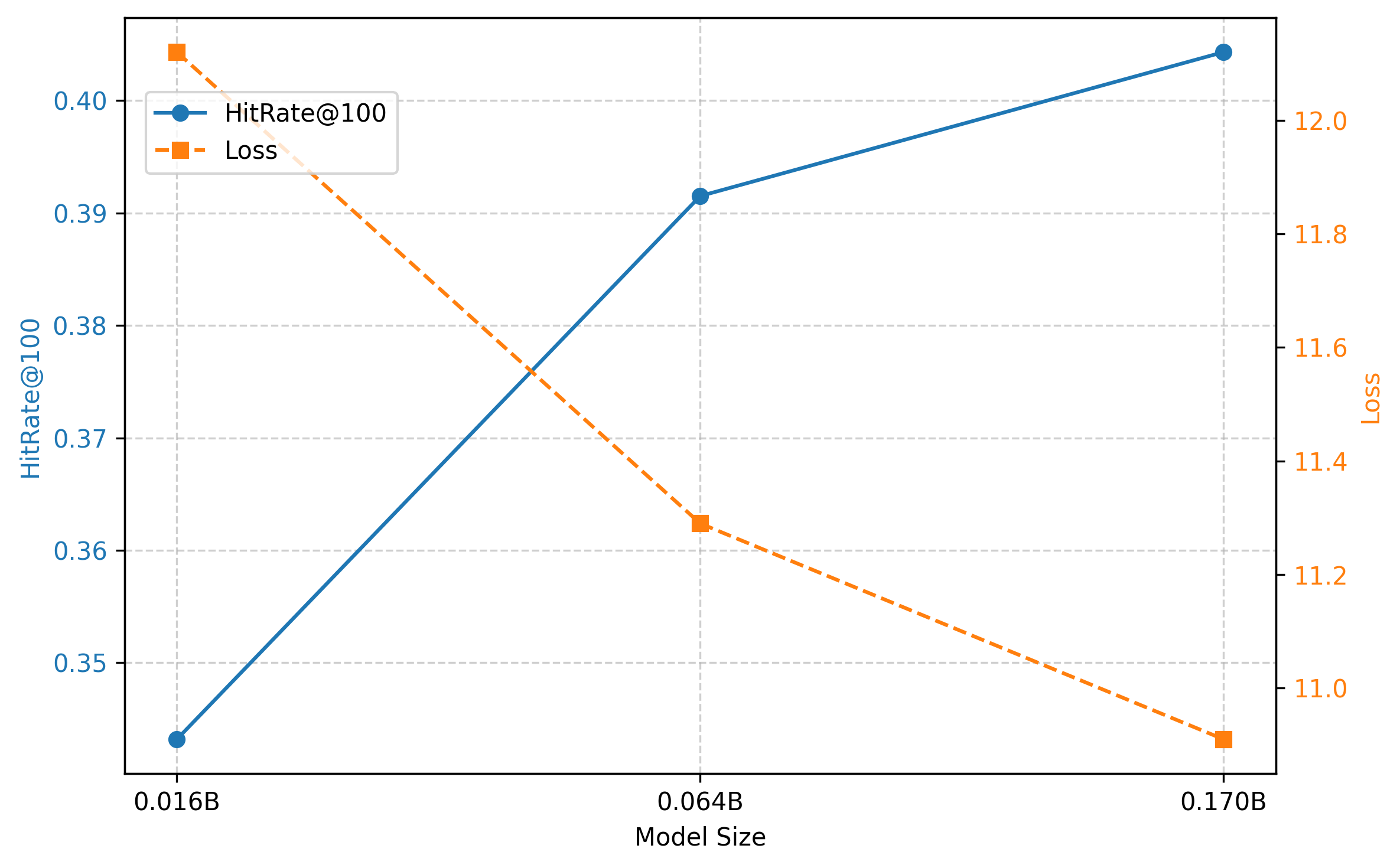}
  \caption{Offline performance comparison of different model sizes.}
  \label{tab:scaling}
\end{figure}

\subsection{Scaling Law Analysis}
\label{sec:scalingLawAnalysis}

We evaluate HGenPush's performance across different model sizes and report HitRate@100 on the video recommendation task as the primary effectiveness metric, while the author recommendation branch exhibits consistent trends.

Specifically, we scale the model through two stages: first, we increase the size from 0.016B to 0.064B by expanding the number of decoder layers from $8$ to $16$ and hidden dimensions from $256$ to $512$. Second, we further scale from 0.064B to 0.170B by replacing dense feed-forward networks with sparse Mixture-of-Experts (MoE) modules \cite{dai2024deepseekmoe, du2022glam}, which efficiently increases model capacity with manageable computational overhead.

As shown in Figure~\ref{tab:scaling}, HGenPush demonstrates favorable scaling properties across the full range. The HitRate@100 metric improves consistently from 0.3432 (0.016B) to 0.3915 (0.064B) to 0.4043 (0.170B). Notably, the first scaling stage achieves a substantial 14.07\% relative improvement, while the second stage yields a more modest 3.27\% gain, reflecting the law of diminishing returns. These results validate that HGenPush can effectively leverage increased model capacity for improved recommendation quality.

\subsection{Reinforcement Learning Performance}
\label{sec:rlPerformance}

While click behaviors reflect users' initial interest, they cannot fully capture content quality and user satisfaction. To address this, we propose introducing real user feedback to help the model identify and boost high-value content, thereby enhancing user experience and the health of the system ecosystem. We compare different methods for incorporating user consumption signals, with offline results shown in Table~\ref{tab:rlOff}. 

\begin{table}[htbp]
  \centering
  \caption{Offline performance comparison of different methods for incorporating user consumption signals.}
  \label{tab:rlOff}
  \begin{tabular}{lcc}
    \toprule
    Training Strategy & HitRate@100 & Loss\\
    \midrule
    Base (Click Labels Only) & 0.3964 & 11.22 \\
    Base + Sample Reweighting & 0.3394 & 12.31 \\
    Base + UCPA & 0.3915 & 11.29 \\
    \bottomrule
  \end{tabular}
\end{table}

 First, we explore sample reweighting based on user watch time, aiming to boost videos with higher user engagement. However, this direct reweighting strategy severely impacts the primary training objective, causing substantial HitRate@100 degradation.
 
 In contrast, our user consumption preference alignment (UCPA) module takes a different method: it groups users' video consumption behaviors within the session following a push click, then computes relative advantage scores for reinforcement learning. This design incorporates much richer consumption signals while maintaining near-baseline performance with only minimal HitRate@100 decline. To further validate the effectiveness of UCPA, we conduct online A/B testing.

Table~\ref{tab:rlOn} shows the improvements in consumption metrics when adding the UCPA module. It can be observed that UCPA achieves consistent improvements across all consumption metrics, demonstrating its effectiveness in promoting high-value content. Notably, Forward Rate increases by 7.57\%, indicating that users are more willing to share recommended content with others. Complete Play Rate and Like Rate also show substantial gains, validating that UCPA successfully guides the model toward content with higher user satisfaction and long-term value.

\begin{table}
    \caption{Online consumption metrics improvement when adding UCPA module compared to click-only baseline.}
    \label{tab:rlOn}
    \begin{tabular}{lcccccc}
    \toprule
    Metric & Improvement \\
    \midrule
    Push Play Duration & +0.43\% \\
    Valid Play Rate & +0.37\% \\
    Long Play Rate & +0.57\% \\
    Complete Play Rate & +0.90\% \\
    Like Rate & +0.75\% \\
    Forward Rate & +7.57\% \\
    \bottomrule
\end{tabular}
\end{table}

\subsection{Online A/B Testing}
\label{sec:onlineABTesting}

To comprehensively evaluate HGenPush's real-world effectiveness, we conduct 
rigorous online A/B testing against the online multi-stage cascading architecture of Kuaishou's push notification system. We primarily report results for the following core metrics:
\begin{itemize}
    \item \textbf{DAU} indicates daily active users.
    \item \textbf{CTR} is user click-through rate for all notifications received in one day.
    \item \textbf{App Usage Duration} is the total time users spend on the application per day.
\end{itemize}

\begin{table*}[htbp]
  \centering
  \caption{Online performance comparison of different HGenPush variants. In parentheses are p-values.}
  \label{tab:online}
  \begin{tabular}{lccc}
    \toprule
    Model Variant & DAU & CTR & App Usage Duration\\
    \midrule
    (v1) Video Branch w/ Chained-MTP & +0.126\% (0.00) & +0.998\% (0.00) & +0.095\% (0.01) \\
    (v2) v1 + UCPA & +0.153\% (0.00) & +1.075\% (0.02) & +0.145\% (0.05) \\
    (v3) HGenPush (v2 + Author Branch) & +0.181\% (0.04) & +1.577\% (0.11) & +0.148\% (0.75) \\
    \bottomrule
  \end{tabular}
\end{table*}

The results in Table~\ref{tab:online} demonstrate consistent improvements through iterative enhancements. First, v1 significantly outperforms the production system across all metrics (+0.126\% DAU, +0.998\% CTR, +0.095\% App Usage Duration), validating that our video recommendation branch can accurately capture user interest in video content. Building upon v1, v2 integrates the UCPA module, yielding further gains. This enhancement stems from UCPA leveraging user consumption signals to strengthen alignment between our generative recommendation and the downstream ranking module, which enables more effective identification of high-quality candidates and ultimately improves experience and engagement. The complete HGenPush (v3) further integrates the author recommendation branch for heterogeneous generation and achieves additional improvements, demonstrating its effectiveness in discovering user trust in authors, thereby satisfying users' follow preference.

In summary, the complete HGenPush achieves +0.181\% DAU, +1.577\% CTR, and +0.148\% App Usage Duration over the production system, validating the effectiveness of our approach. Given the scale of our platform and the maturity of the production baseline, these improvements demonstrate substantial business value and prove the practical viability of HGenPush in large-scale production systems.

\section{Conclusion}

Existing generative recommendation systems struggle with heterogeneous recommendation and suffer from low inference efficiency due to autoregressive generation. To address this, we propose HGenPush, a heterogeneous generative recommendation architecture for industrial push notification systems. The key contributions of our approach are as follows: First, we achieve joint recommendation of heterogeneous content (videos and authors) in the generative recommendation paradigm. Second, we innovatively propose a multi-token prediction method called Chained-MTP, which utilizes user interest representation and cumulative semantic embedding to prevent interest drift and preserve semantic dependencies, while enhancing inference efficiency. Additionally, we design a user consumption preference alignment module to encourage the recommendation of content with long-term value. Experimental results demonstrate that HGenPush achieves significant improvements in both effectiveness and efficiency for real-world deployment.


\bibliographystyle{ACM-Reference-Format}
\balance
\bibliography{arxiv}


\begin{thebibliography}{24}


\ifx \showCODEN    \undefined \def \showCODEN     #1{\unskip}     \fi
\ifx \showISBNx    \undefined \def \showISBNx     #1{\unskip}     \fi
\ifx \showISBNxiii \undefined \def \showISBNxiii  #1{\unskip}     \fi
\ifx \showISSN     \undefined \def \showISSN      #1{\unskip}     \fi
\ifx \showLCCN     \undefined \def \showLCCN      #1{\unskip}     \fi
\ifx \shownote     \undefined \def \shownote      #1{#1}          \fi
\ifx \showarticletitle \undefined \def \showarticletitle #1{#1}   \fi
\ifx \showURL      \undefined \def \showURL       {\relax}        \fi
\providecommand\bibfield[2]{#2}
\providecommand\bibinfo[2]{#2}
\providecommand\natexlab[1]{#1}
\providecommand\showeprint[2][]{arXiv:#2}

\bibitem[Chen et~al\mbox{.}(2025)]%
        {chen2025minimax}
\bibfield{author}{\bibinfo{person}{Aili Chen}, \bibinfo{person}{Aonian Li}, \bibinfo{person}{Bangwei Gong}, \bibinfo{person}{Binyang Jiang}, \bibinfo{person}{Bo Fei}, \bibinfo{person}{Bo Yang}, \bibinfo{person}{Boji Shan}, \bibinfo{person}{Changqing Yu}, \bibinfo{person}{Chao Wang}, \bibinfo{person}{Cheng Zhu}, {et~al\mbox{.}}} \bibinfo{year}{2025}\natexlab{}.
\newblock \showarticletitle{MiniMax-M1: Scaling Test-Time Compute Efficiently with Lightning Attention}.
\newblock \bibinfo{journal}{\emph{arXiv preprint arXiv:2506.13585}} (\bibinfo{year}{2025}).
\newblock


\bibitem[Dai et~al\mbox{.}(2024)]%
        {dai2024deepseekmoe}
\bibfield{author}{\bibinfo{person}{Damai Dai}, \bibinfo{person}{Chengqi Deng}, \bibinfo{person}{Chenggang Zhao}, \bibinfo{person}{RX Xu}, \bibinfo{person}{Huazuo Gao}, \bibinfo{person}{Deli Chen}, \bibinfo{person}{Jiashi Li}, \bibinfo{person}{Wangding Zeng}, \bibinfo{person}{Xingkai Yu}, \bibinfo{person}{Yu Wu}, {et~al\mbox{.}}} \bibinfo{year}{2024}\natexlab{}.
\newblock \showarticletitle{Deepseekmoe: Towards ultimate expert specialization in mixture-of-experts language models}. In \bibinfo{booktitle}{\emph{Proceedings of the 62nd Annual Meeting of the Association for Computational Linguistics (Volume 1: Long Papers)}}. \bibinfo{pages}{1280--1297}.
\newblock


\bibitem[Deng et~al\mbox{.}(2025)]%
        {deng2025onerec}
\bibfield{author}{\bibinfo{person}{Jiaxin Deng}, \bibinfo{person}{Shiyao Wang}, \bibinfo{person}{Kuo Cai}, \bibinfo{person}{Lejian Ren}, \bibinfo{person}{Qigen Hu}, \bibinfo{person}{Weifeng Ding}, \bibinfo{person}{Qiang Luo}, {and} \bibinfo{person}{Guorui Zhou}.} \bibinfo{year}{2025}\natexlab{}.
\newblock \showarticletitle{Onerec: Unifying retrieve and rank with generative recommender and iterative preference alignment}.
\newblock \bibinfo{journal}{\emph{arXiv preprint arXiv:2502.18965}} (\bibinfo{year}{2025}).
\newblock


\bibitem[Du et~al\mbox{.}(2022)]%
        {du2022glam}
\bibfield{author}{\bibinfo{person}{Nan Du}, \bibinfo{person}{Yanping Huang}, \bibinfo{person}{Andrew~M Dai}, \bibinfo{person}{Simon Tong}, \bibinfo{person}{Dmitry Lepikhin}, \bibinfo{person}{Yuanzhong Xu}, \bibinfo{person}{Maxim Krikun}, \bibinfo{person}{Yanqi Zhou}, \bibinfo{person}{Adams~Wei Yu}, \bibinfo{person}{Orhan Firat}, {et~al\mbox{.}}} \bibinfo{year}{2022}\natexlab{}.
\newblock \showarticletitle{Glam: Efficient scaling of language models with mixture-of-experts}. In \bibinfo{booktitle}{\emph{International conference on machine learning}}. PMLR, \bibinfo{pages}{5547--5569}.
\newblock


\bibitem[Guo et~al\mbox{.}(2025)]%
        {guo2025onesug}
\bibfield{author}{\bibinfo{person}{Xian Guo}, \bibinfo{person}{Ben Chen}, \bibinfo{person}{Siyuan Wang}, \bibinfo{person}{Ying Yang}, \bibinfo{person}{Chenyi Lei}, \bibinfo{person}{Yuqing Ding}, {and} \bibinfo{person}{Han Li}.} \bibinfo{year}{2025}\natexlab{}.
\newblock \showarticletitle{OneSug: The Unified End-to-End Generative Framework for E-commerce Query Suggestion}.
\newblock \bibinfo{journal}{\emph{arXiv preprint arXiv:2506.06913}} (\bibinfo{year}{2025}).
\newblock


\bibitem[Kang and McAuley(2018)]%
        {kang2018self}
\bibfield{author}{\bibinfo{person}{Wang-Cheng Kang} {and} \bibinfo{person}{Julian McAuley}.} \bibinfo{year}{2018}\natexlab{}.
\newblock \showarticletitle{Self-attentive sequential recommendation}. In \bibinfo{booktitle}{\emph{2018 IEEE international conference on data mining (ICDM)}}. IEEE, \bibinfo{pages}{197--206}.
\newblock


\bibitem[Lambert et~al\mbox{.}(2024)]%
        {lambert2024tulu}
\bibfield{author}{\bibinfo{person}{Nathan Lambert}, \bibinfo{person}{Jacob Morrison}, \bibinfo{person}{Valentina Pyatkin}, \bibinfo{person}{Shengyi Huang}, \bibinfo{person}{Hamish Ivison}, \bibinfo{person}{Faeze Brahman}, \bibinfo{person}{Lester James~V Miranda}, \bibinfo{person}{Alisa Liu}, \bibinfo{person}{Nouha Dziri}, \bibinfo{person}{Shane Lyu}, {et~al\mbox{.}}} \bibinfo{year}{2024}\natexlab{}.
\newblock \showarticletitle{Tulu 3: Pushing frontiers in open language model post-training}.
\newblock \bibinfo{journal}{\emph{arXiv preprint arXiv:2411.15124}} (\bibinfo{year}{2024}).
\newblock


\bibitem[Li et~al\mbox{.}(2023)]%
        {li2023text}
\bibfield{author}{\bibinfo{person}{Jiacheng Li}, \bibinfo{person}{Ming Wang}, \bibinfo{person}{Jin Li}, \bibinfo{person}{Jinmiao Fu}, \bibinfo{person}{Xin Shen}, \bibinfo{person}{Jingbo Shang}, {and} \bibinfo{person}{Julian McAuley}.} \bibinfo{year}{2023}\natexlab{}.
\newblock \showarticletitle{Text is all you need: Learning language representations for sequential recommendation}. In \bibinfo{booktitle}{\emph{Proceedings of the 29th ACM SIGKDD Conference on Knowledge Discovery and Data Mining}}. \bibinfo{pages}{1258--1267}.
\newblock


\bibitem[Li et~al\mbox{.}(2024)]%
        {li2024calrec}
\bibfield{author}{\bibinfo{person}{Yaoyiran Li}, \bibinfo{person}{Xiang Zhai}, \bibinfo{person}{Moustafa Alzantot}, \bibinfo{person}{Keyi Yu}, \bibinfo{person}{Ivan Vuli{\'c}}, \bibinfo{person}{Anna Korhonen}, {and} \bibinfo{person}{Mohamed Hammad}.} \bibinfo{year}{2024}\natexlab{}.
\newblock \showarticletitle{Calrec: Contrastive alignment of generative llms for sequential recommendation}. In \bibinfo{booktitle}{\emph{Proceedings of the 18th ACM Conference on Recommender Systems}}. \bibinfo{pages}{422--432}.
\newblock


\bibitem[Liu et~al\mbox{.}(2024)]%
        {liu2024deepseek}
\bibfield{author}{\bibinfo{person}{Aixin Liu}, \bibinfo{person}{Bei Feng}, \bibinfo{person}{Bing Xue}, \bibinfo{person}{Bingxuan Wang}, \bibinfo{person}{Bochao Wu}, \bibinfo{person}{Chengda Lu}, \bibinfo{person}{Chenggang Zhao}, \bibinfo{person}{Chengqi Deng}, \bibinfo{person}{Chenyu Zhang}, \bibinfo{person}{Chong Ruan}, {et~al\mbox{.}}} \bibinfo{year}{2024}\natexlab{}.
\newblock \showarticletitle{Deepseek-v3 technical report}.
\newblock \bibinfo{journal}{\emph{arXiv preprint arXiv:2412.19437}} (\bibinfo{year}{2024}).
\newblock


\bibitem[Luo et~al\mbox{.}(2025)]%
        {luo2025qarm}
\bibfield{author}{\bibinfo{person}{Xinchen Luo}, \bibinfo{person}{Jiangxia Cao}, \bibinfo{person}{Tianyu Sun}, \bibinfo{person}{Jinkai Yu}, \bibinfo{person}{Rui Huang}, \bibinfo{person}{Wei Yuan}, \bibinfo{person}{Hezheng Lin}, \bibinfo{person}{Yichen Zheng}, \bibinfo{person}{Shiyao Wang}, \bibinfo{person}{Qigen Hu}, {et~al\mbox{.}}} \bibinfo{year}{2025}\natexlab{}.
\newblock \showarticletitle{Qarm: Quantitative alignment multi-modal recommendation at kuaishou}. In \bibinfo{booktitle}{\emph{Proceedings of the 34th ACM International Conference on Information and Knowledge Management}}. \bibinfo{pages}{5915--5922}.
\newblock


\bibitem[Rajput et~al\mbox{.}(2023)]%
        {rajput2023recommender}
\bibfield{author}{\bibinfo{person}{Shashank Rajput}, \bibinfo{person}{Nikhil Mehta}, \bibinfo{person}{Anima Singh}, \bibinfo{person}{Raghunandan Hulikal~Keshavan}, \bibinfo{person}{Trung Vu}, \bibinfo{person}{Lukasz Heldt}, \bibinfo{person}{Lichan Hong}, \bibinfo{person}{Yi Tay}, \bibinfo{person}{Vinh Tran}, \bibinfo{person}{Jonah Samost}, {et~al\mbox{.}}} \bibinfo{year}{2023}\natexlab{}.
\newblock \showarticletitle{Recommender systems with generative retrieval}.
\newblock \bibinfo{journal}{\emph{Advances in Neural Information Processing Systems}}  \bibinfo{volume}{36} (\bibinfo{year}{2023}), \bibinfo{pages}{10299--10315}.
\newblock


\bibitem[Schulman et~al\mbox{.}(2017)]%
        {schulman2017proximal}
\bibfield{author}{\bibinfo{person}{John Schulman}, \bibinfo{person}{Filip Wolski}, \bibinfo{person}{Prafulla Dhariwal}, \bibinfo{person}{Alec Radford}, {and} \bibinfo{person}{Oleg Klimov}.} \bibinfo{year}{2017}\natexlab{}.
\newblock \showarticletitle{Proximal policy optimization algorithms}.
\newblock \bibinfo{journal}{\emph{arXiv preprint arXiv:1707.06347}} (\bibinfo{year}{2017}).
\newblock


\bibitem[Shao et~al\mbox{.}(2024)]%
        {shao2024deepseekmath}
\bibfield{author}{\bibinfo{person}{Zhihong Shao}, \bibinfo{person}{Peiyi Wang}, \bibinfo{person}{Qihao Zhu}, \bibinfo{person}{Runxin Xu}, \bibinfo{person}{Junxiao Song}, \bibinfo{person}{Xiao Bi}, \bibinfo{person}{Haowei Zhang}, \bibinfo{person}{Mingchuan Zhang}, \bibinfo{person}{YK Li}, \bibinfo{person}{Yang Wu}, {et~al\mbox{.}}} \bibinfo{year}{2024}\natexlab{}.
\newblock \showarticletitle{Deepseekmath: Pushing the limits of mathematical reasoning in open language models}.
\newblock \bibinfo{journal}{\emph{arXiv preprint arXiv:2402.03300}} (\bibinfo{year}{2024}).
\newblock


\bibitem[Shazeer(2020)]%
        {shazeer2020glu}
\bibfield{author}{\bibinfo{person}{Noam Shazeer}.} \bibinfo{year}{2020}\natexlab{}.
\newblock \showarticletitle{Glu variants improve transformer}.
\newblock \bibinfo{journal}{\emph{arXiv preprint arXiv:2002.05202}} (\bibinfo{year}{2020}).
\newblock


\bibitem[Wang et~al\mbox{.}(2025)]%
        {wang2025generative}
\bibfield{author}{\bibinfo{person}{Dongsheng Wang}, \bibinfo{person}{Yuxi Huang}, \bibinfo{person}{Shen Gao}, \bibinfo{person}{Yifan Wang}, \bibinfo{person}{Chengrui Huang}, {and} \bibinfo{person}{Shuo Shang}.} \bibinfo{year}{2025}\natexlab{}.
\newblock \showarticletitle{Generative next poi recommendation with semantic id}. In \bibinfo{booktitle}{\emph{Proceedings of the 31st ACM SIGKDD Conference on Knowledge Discovery and Data Mining V. 2}}. \bibinfo{pages}{2904--2914}.
\newblock


\bibitem[Wang et~al\mbox{.}(2024)]%
        {wang2024enhanced}
\bibfield{author}{\bibinfo{person}{Yidan Wang}, \bibinfo{person}{Zhaochun Ren}, \bibinfo{person}{Weiwei Sun}, \bibinfo{person}{Jiyuan Yang}, \bibinfo{person}{Zhixiang Liang}, \bibinfo{person}{Xin Chen}, \bibinfo{person}{Ruobing Xie}, \bibinfo{person}{Su Yan}, \bibinfo{person}{Xu Zhang}, \bibinfo{person}{Pengjie Ren}, {et~al\mbox{.}}} \bibinfo{year}{2024}\natexlab{}.
\newblock \showarticletitle{Enhanced generative recommendation via content and collaboration integration}.
\newblock \bibinfo{journal}{\emph{CoRR}} (\bibinfo{year}{2024}).
\newblock


\bibitem[Wei et~al\mbox{.}(2025)]%
        {wei2025oneloc}
\bibfield{author}{\bibinfo{person}{Zhipeng Wei}, \bibinfo{person}{Kuo Cai}, \bibinfo{person}{Junda She}, \bibinfo{person}{Jie Chen}, \bibinfo{person}{Minghao Chen}, \bibinfo{person}{Yang Zeng}, \bibinfo{person}{Qiang Luo}, \bibinfo{person}{Wencong Zeng}, \bibinfo{person}{Ruiming Tang}, \bibinfo{person}{Kun Gai}, {et~al\mbox{.}}} \bibinfo{year}{2025}\natexlab{}.
\newblock \showarticletitle{Oneloc: Geo-aware generative recommender systems for local life service}.
\newblock \bibinfo{journal}{\emph{arXiv preprint arXiv:2508.14646}} (\bibinfo{year}{2025}).
\newblock


\bibitem[Zhai et~al\mbox{.}(2024)]%
        {zhai2024actions}
\bibfield{author}{\bibinfo{person}{Jiaqi Zhai}, \bibinfo{person}{Lucy Liao}, \bibinfo{person}{Xing Liu}, \bibinfo{person}{Yueming Wang}, \bibinfo{person}{Rui Li}, \bibinfo{person}{Xuan Cao}, \bibinfo{person}{Leon Gao}, \bibinfo{person}{Zhaojie Gong}, \bibinfo{person}{Fangda Gu}, \bibinfo{person}{Michael He}, {et~al\mbox{.}}} \bibinfo{year}{2024}\natexlab{}.
\newblock \showarticletitle{Actions speak louder than words: Trillion-parameter sequential transducers for generative recommendations}.
\newblock \bibinfo{journal}{\emph{arXiv preprint arXiv:2402.17152}} (\bibinfo{year}{2024}).
\newblock


\bibitem[Zhang and Sennrich(2019)]%
        {zhang2019root}
\bibfield{author}{\bibinfo{person}{Biao Zhang} {and} \bibinfo{person}{Rico Sennrich}.} \bibinfo{year}{2019}\natexlab{}.
\newblock \showarticletitle{Root mean square layer normalization}.
\newblock \bibinfo{journal}{\emph{Advances in neural information processing systems}}  \bibinfo{volume}{32} (\bibinfo{year}{2019}).
\newblock


\bibitem[Zheng et~al\mbox{.}(2024)]%
        {zheng2024adapting}
\bibfield{author}{\bibinfo{person}{Bowen Zheng}, \bibinfo{person}{Yupeng Hou}, \bibinfo{person}{Hongyu Lu}, \bibinfo{person}{Yu Chen}, \bibinfo{person}{Wayne~Xin Zhao}, \bibinfo{person}{Ming Chen}, {and} \bibinfo{person}{Ji-Rong Wen}.} \bibinfo{year}{2024}\natexlab{}.
\newblock \showarticletitle{Adapting large language models by integrating collaborative semantics for recommendation}. In \bibinfo{booktitle}{\emph{2024 IEEE 40th International Conference on Data Engineering (ICDE)}}. IEEE, \bibinfo{pages}{1435--1448}.
\newblock


\bibitem[Zheng et~al\mbox{.}(2025a)]%
        {zheng2025group}
\bibfield{author}{\bibinfo{person}{Chujie Zheng}, \bibinfo{person}{Shixuan Liu}, \bibinfo{person}{Mingze Li}, \bibinfo{person}{Xiong-Hui Chen}, \bibinfo{person}{Bowen Yu}, \bibinfo{person}{Chang Gao}, \bibinfo{person}{Kai Dang}, \bibinfo{person}{Yuqiong Liu}, \bibinfo{person}{Rui Men}, \bibinfo{person}{An Yang}, {et~al\mbox{.}}} \bibinfo{year}{2025}\natexlab{a}.
\newblock \showarticletitle{Group sequence policy optimization}.
\newblock \bibinfo{journal}{\emph{arXiv preprint arXiv:2507.18071}} (\bibinfo{year}{2025}).
\newblock


\bibitem[Zheng et~al\mbox{.}(2025b)]%
        {zheng2025ega}
\bibfield{author}{\bibinfo{person}{Zuowu Zheng}, \bibinfo{person}{Ze Wang}, \bibinfo{person}{Fan Yang}, \bibinfo{person}{Jiangke Fan}, \bibinfo{person}{Teng Zhang}, {and} \bibinfo{person}{Xingxing Wang}.} \bibinfo{year}{2025}\natexlab{b}.
\newblock \showarticletitle{EGA: A Unified End-to-End Generative Framework for Industrial Advertising Systems}.
\newblock \bibinfo{journal}{\emph{arXiv preprint arXiv:2505.17549}} (\bibinfo{year}{2025}).
\newblock


\bibitem[Zhu et~al\mbox{.}(2025)]%
        {zhu2025addressing}
\bibfield{author}{\bibinfo{person}{Yongxin Zhu}, \bibinfo{person}{Bocheng Li}, \bibinfo{person}{Yifei Xin}, \bibinfo{person}{Zhihua Xia}, {and} \bibinfo{person}{Linli Xu}.} \bibinfo{year}{2025}\natexlab{}.
\newblock \showarticletitle{Addressing representation collapse in vector quantized models with one linear layer}. In \bibinfo{booktitle}{\emph{Proceedings of the IEEE/CVF International Conference on Computer Vision}}. \bibinfo{pages}{22968--22977}.
\newblock


\end{thebibliography}


\end{document}